\documentclass[footinbib,aps,pra,reprint,superscriptaddress,nopacs]{revtex4-1}

\usepackage{graphicx}
\usepackage{amsmath}
\usepackage{amsfonts}
\usepackage{verbatim}
\usepackage[absolute]{textpos}
\usepackage{ragged2e}

\DeclareMathOperator{\Tr}{Tr}

\begin{document}
\title{Electro-Optic Frequency Beamsplitters and Tritters for High-Fidelity Photonic Quantum Information Processing}

\author{Hsuan-Hao Lu}
\affiliation{School of Electrical and Computer Engineering and Purdue Quantum Center, Purdue University, West Lafayette, Indiana 47907, USA}
\author{Joseph M. Lukens}
\affiliation{Quantum Information Science Group, Computational Sciences and Engineering Division, Oak Ridge National Laboratory, Oak Ridge, Tennessee 37831, USA}
\author{Nicholas A. Peters}
\affiliation{Quantum Information Science Group, Computational Sciences and Engineering Division, Oak Ridge National Laboratory, Oak Ridge, Tennessee 37831, USA}
\affiliation{Bredesen Center for Interdisciplinary Research and Graduate Education, The University of Tennessee, Knoxville, Tennessee 37996, USA}
\author{Ogaga D. Odele}
\author{Daniel E. Leaird}
\author{Andrew M. Weiner}
\affiliation{School of Electrical and Computer Engineering and Purdue Quantum Center, Purdue University, West Lafayette, Indiana 47907, USA}
\author{Pavel Lougovski}
\email{lougovskip@ornl.gov}
\affiliation{Quantum Information Science Group, Computational Sciences and Engineering Division, Oak Ridge National Laboratory, Oak Ridge, Tennessee 37831, USA}

\date{\today}

\begin{abstract}
We report experimental realization of high-fidelity photonic quantum gates for frequency-encoded qubits and qutrits based on electro-optic modulation and Fourier-transform pulse shaping. Our frequency version of the Hadamard gate offers near-unity fidelity ($0.99998\pm0.00003$), requires only a single microwave drive tone for near-ideal performance, functions across the entire C-band (1530-1570 nm), and can operate concurrently on multiple qubits spaced as tightly as four frequency modes apart, with no observable degradation in the fidelity. For qutrits we implement a $3\times 3$ extension of the Hadamard gate: the balanced tritter. This tritter---the first ever demonstrated for frequency modes---attains fidelity $0.9989\pm0.0004$. These gates represent important building blocks toward scalable, high-fidelity quantum information processing based on frequency encoding.
\end{abstract}

\maketitle

\textit{Introduction.---}The coherent translation of quantum states from one frequency to another via optical nonlinearites has been the focus of considerable research since the early 1990s~\cite{Kumar1992}; yet only fairly recently have such processes been explored in the more elaborate context of time-frequency quantum information processing (QIP), where optical frequency is not just the \emph{carrier} of quantum information but the \emph{information itself}. Important examples include the quantum pulse gate~\cite{Brecht2011, Eckstein2011}, which uses nonlinear mixing with shaped classical pulses for selective conversion of the time-frequency modes of single photons~\cite{Brecht2014, Manurkar2016, Ansari2016}, and demonstrations of frequency beamsplitters based on both $\chi^{(2)}$~\cite{Kobayashi2016, Kobayashi2017} and $\chi^{(3)}$~\cite{McGuinness2010, Clemmen2016, Joshi2017} nonlinearities, which interfere two wavelength modes analogously to a spatial beamsplitter. These seminal experiments have shown key primitives in frequency-based QIP, but many challenges remain. For example, optical filters and/or low temperatures are required to remove background noise due to powerful optical pumps, either from  the sources themselves or Raman scattering in the nonlinear medium. And achieving the necessary nonlinear mixing for arbitrary combinations of modes will require additional pump fields, as well as properly engineered phase-matching conditions. 

Recently we proposed a fundamentally distinct platform for frequency-bin manipulations, relying on electro-optic phase modulation and Fourier-transform pulse shaping for universal QIP~\cite{Lukens2017}. Our approach requires no optical pump fields, is readily parallelized, and scales well with the number of modes. In this Letter, we apply this paradigm to experimentally demonstrate the first electro-optic-based frequency beamsplitter. Our frequency beamsplitter attains high fidelity, operates in parallel on multiple two-mode subsets across the entire optical C-band, and retains excellent performance at the single-photon level. Moreover, by incorporating an additional harmonic in the microwave drive signal, we also realize a balanced frequency tritter, the three-mode extension of the beamsplitter. This is the first frequency tritter demonstrated on any platform, and establishes our electro-optic approach as a leader for high-dimensional frequency-based QIP. Combined with its native parallelizability and absence of optical noise sources, our mixer design offers new opportunities for a range of quantum information applications, including linear-optical computation~\cite{Lukens2017}, quantum repeaters~\cite{azuma2015all}, and quantum walks~\cite{hillery2003quantum}. The tritter also serves as an elementary building block for a frequency version of three-mode directionally unbiased linear-optical multiports, which find application in quantum simulations~\cite{simon2017quantum} and Bell state discriminators~\cite{simon2016group}. 

\textit{Background.---}The Hilbert space of interest consists of a comb of equispaced frequency bins, with operators $\hat{a}_n$ ($n \in \mathbb{Z}$) that annihilate a single photon in the narrowband modes centered at frequencies $\omega_n=\omega_0+n\Delta\omega$~\cite{Olislager2010, Lukens2017}. A qudit is represented by a single photon spread over $d$ such modes, and the objective is to implement a frequency multiport $V$ connecting the input $\hat{a}_n^{(\mathrm{in})}$ and output $\hat{a}_m^{(\mathrm{out})}$ modes in some desired fashion: $\hat{a}_m^{(\mathrm{out})} = \sum_n V_{mn} \hat{a}_n^{(\mathrm{in})}$. Line-by-line pulse shaping~\cite{Cundiff2010, Weiner2011} permits arbitrary phase shifts for frequency modes, i.e., the operation $V_{mn}=e^{i\phi_m} \delta_{mn}$. Following the initial demonstration of entangled-photon temporal shaping in 2005~\cite{Peer2005}, a range of experiments have showcased the utility of pulse shaping at the single-photon level~\cite{Zaeh2008, Bernhard2013, Lukens2013c, Lukens2014a, Agarwal2014}.

However, universal QIP also requires frequency mode \emph{mixing}. And while, as noted above, parametric processes have enabled two-mode frequency beamsplitters, electro-optic modulation represents an attractive alternative: it requires no optical pumps, relies on purely electrical controls, and is compatible with state-of-the-art telecommunication technology. Such features have enabled impressive electro-optic experiments in quantum photonics, including single-photon temporal shaping~\cite{Kolchin2008, Belthangady2010, Liu2014, Karpinski2017, Wright2017}, nonlocal modulation cancellation~\cite{Harris2008, Sensarn2009a, Olislager2010}, and state measurement~\cite{Belthangady2009, Lukens2015}. Nevertheless, realization of an arbitrary $d\times d$ frequency-bin multiport presents stark challenges for a single electro-optic phase modulator (EOM). By design, an EOM couples a single input frequency mode to many output modes, unavoidably scattering an input photon outside of the $d$-dimensional computational space. A simple argument suggests that this undesired ``scatter probability'' is at least $(d-1)/(2d-1)$ for a uniform $d$-mode mixer based on a single EOM ~(Appendix~\ref{appA}). Yet this limitation can be circumvented by considering two EOMs with a pulse shaper sandwiched between them; the spectral phase imparted by the middle stage ensures that the sidebands populated after the first EOM are returned to the computational space after the second one, thereby making it possible to realize a fully deterministic frequency beamsplitter~\cite{Lukens2017}.

Quantitatively, the performance of a generic frequency multiport $V$ can be compared to the desired $d\times d$ unitary operation $U_{d\times d}$ through success probability $\mathcal{P} = \frac{1}{d} \Tr (V_{d\times d}^\dagger V_{d\times d})$ and fidelity $\mathcal{F} = \frac{1}{\mathcal{P}d^{2}} |\Tr (V_{d\times d}^\dagger U_{d\times d})|^2$ metrics, where $V_{d \times d}$ denotes the infinite-dimensional unitary $V$ truncated to the $d$ modes of $U_{d\times d}$~\cite{Uskov2009}. Experimentally, the success probability is further degraded by photon loss, an effect absent in an ideal unitary. But since insertion loss is distinct from operation purity---the former being technical in nature, the latter stemming from fundamental properties of the modulation approach---we normalize the measured linear transformation by total transmissivity before computing $\mathcal{P}$.

\textit{Frequency beamsplitter.---}For our first experimental demonstration, we focus on the 50/50 beamsplitter with phases chosen to match the Hadamard gate:
\begin{equation}
\label{e4.1}
U_{2\times 2} = \frac{1}{\sqrt{2}} \begin{pmatrix} 1 & 1 \\ 1 & -1 \end{pmatrix},
\end{equation}
the top row corresponding to mode 0 ($\omega_0$) and the bottom to mode 1 ($\omega_1$).
We make use of two improvements from our original solution in~\cite{Lukens2017}, which result in a more practical  experimental setup. First, we can absorb the initial pulse shaper into the first EOM, thereby reducing the number of optical components from four to three; second, by considering only phase-shifted sinewaves as the electro-optic modulation functions---rather than arbitrary waveforms---theory still predicts $\mathcal{F}=0.9999$ and $\mathcal{P}=0.9760$~(Appendix~\ref{appB}): a small reduction from unity and well above the single-EOM limit of $\mathcal{P}=2/3$. This near-ideal performance even with such simple microwave modulation represents a major theoretical advance in terms of practicality and scalability, removing the need for a high-bandwidth arbitrary waveform generator to realize the Hadamard gate. Moreover, while we focus on nearest-neighbor mode coupling, in which the microwave drive frequency equals the fundamental mode spacing $\Delta\omega$, spectrally separated modes can be mixed as well. Setting the modulation frequency to an integer multiple $N\Delta\omega$ produces a frequency beasmplitter for lines now spaced $N$ modes apart, all while avoiding crosstalk with interior modes, assuming a pure $N\Delta\omega$-periodic drive. Using the pulse-shaper phases, it is then even possible to realize different operations on these interleaved $N$-harmonic ``supergrids,'' potentially permitting an array of independent nearest-neighbor and nonadjacent frequency operations within the same set of elements.

\begin{figure}[!b]
\includegraphics[width=3.4in]{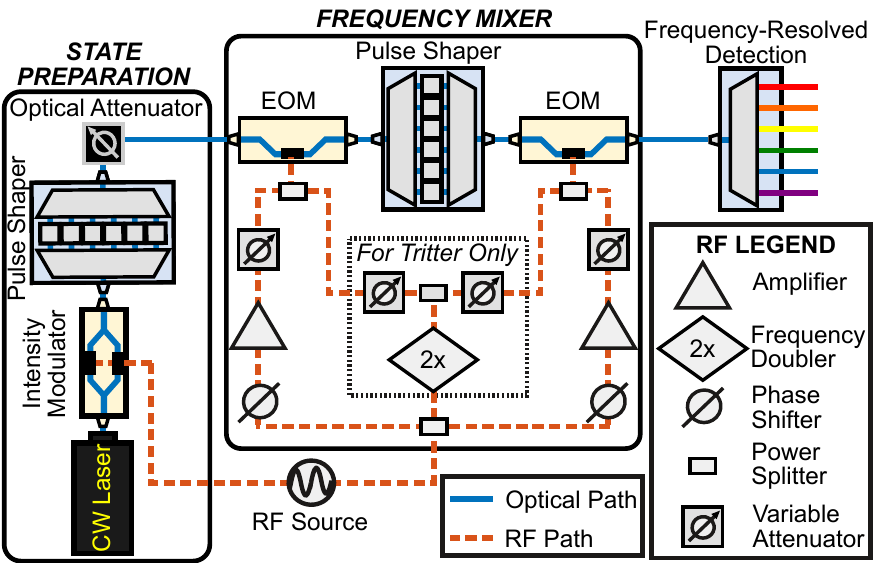}
\caption{Experimental setup. See text and Appendix~\ref{appC} for details.}
\label{fig1}
\end{figure}

Figure~\ref{fig1} provides a schematic of the experimental setup~(Appendix~\ref{appC}). A radio-frequency (RF) oscillator provides a 25-GHz drive signal to each EOM, with amplifiers and delay lines setting the appropriate amplitude and phase for each waveform. The central pulse shaper applies the numerically optimized spectral phase pattern for the Hadamard gate. The $\sim$10-GHz spectral resolution of this pulse shaper ultimately limits the tightest frequency-mode spacing (and thus total number of modes) we can utilize in our setup; experimentally we have found detectable reduction in $\mathcal{F}$ and $\mathcal{P}$ for spacings below $\sim$18 GHz. To characterize the full frequency-bin multiport, we probe it with an electro-optic frequency comb, measuring the output spectrum for different input frequency superpositions. This technique represents the analogue of the spatial version proposed and demonstrated in~\cite{Rahimi2013}, applied here for the first time to frequency modes~(Appendix~\ref{appD}). We also adopt the convention~\cite{Rahimi2013} which specifies zero phase as the input superposition state that maximizes the power in the zeroth frequency bin of the output; the phase values of any subsequent state (as set by the state preparation pulse shaper in Fig.~\ref{fig1}) are thus only defined relative to this operating point. At a center wavelength of 1545.04 nm ($\omega_0 = 2\pi \times 194.036$ THz), we measure fidelity $\mathcal{F} = 0.99998 \pm 0.00003$ and success probability $\mathcal{P} = 0.9739 \pm 0.0003$, where error bars give the standard deviation of five independent measurement sequences. The current gate insertion loss is 12.5 dB: the EOMs contribute $\sim$2.8 dB each; the pulse shaper, $\sim$4.7 dB; and the remainder comes from fiber patch cord connections and polarization controllers.

\begin{figure}[!b]
\includegraphics[width=3.4in]{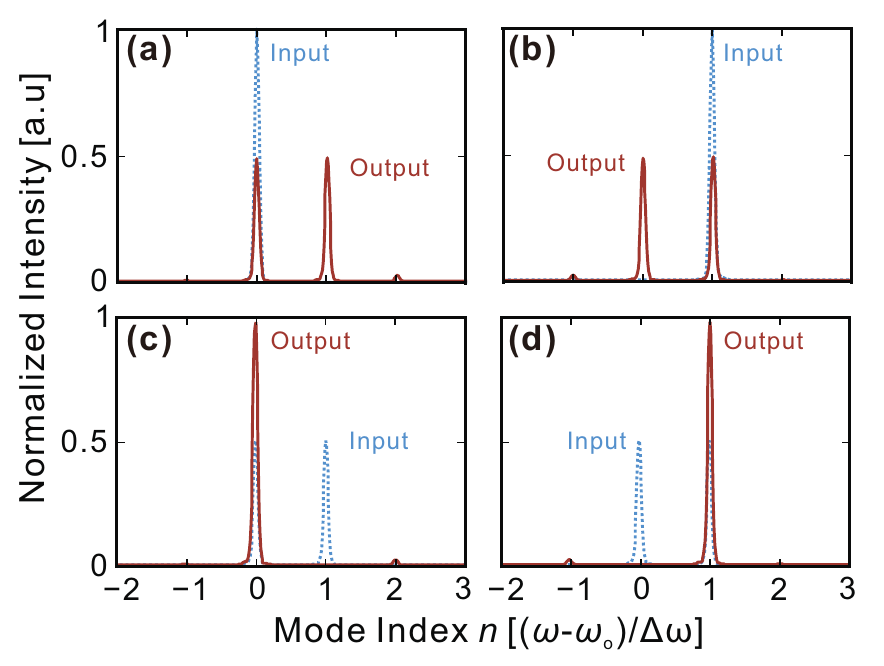}
\caption{Experimentally measured beamsplitter output spectra for specific coherent state inputs. (a) Pure mode 0: $|\alpha_{\omega_0} 0_{\omega_1}\rangle$. (b) Pure mode 1: $|0_{\omega_0} \alpha_{\omega_1}\rangle$. (c) Mode 0 and mode 1 in phase: $|\alpha_{\omega_0} \alpha_{\omega_1}\rangle$. (d) Mode 0 and mode 1 out of phase: $|\alpha_{\omega_0} (-\alpha)_{\omega_1}\rangle$.
}
\label{fig2}
\end{figure}

Figure~\ref{fig2} shows four experimentally recorded input/output combinations: the top row shows the equi-amplitude superpositions resulting from input in either mode 0 or mode 1; the second row reveals the single-wavelength output with the input in the states $|\alpha_{\omega_0} (\pm\alpha)_{\omega_1}\rangle$. The small bumps in adjacent modes $-1$ and $+2$ reflect the nonunity success probability, a limitation which---as noted above---could be removed by more sophisticated modulation waveforms. And even in the current arrangement with $\mathcal{P}\approx 0.97$, the impact such residual scattering could have on gates downstream---i.e., by coupling back into the computational space and introducing errors---can be eliminated, either by using the next pulse shaper to selectively attenuate these modes, or by sending them to a fiber tap for detection.

\begin{figure}
\includegraphics[width=3.4in, trim={0.05in 0.05in 0.05in 0.05in},clip]{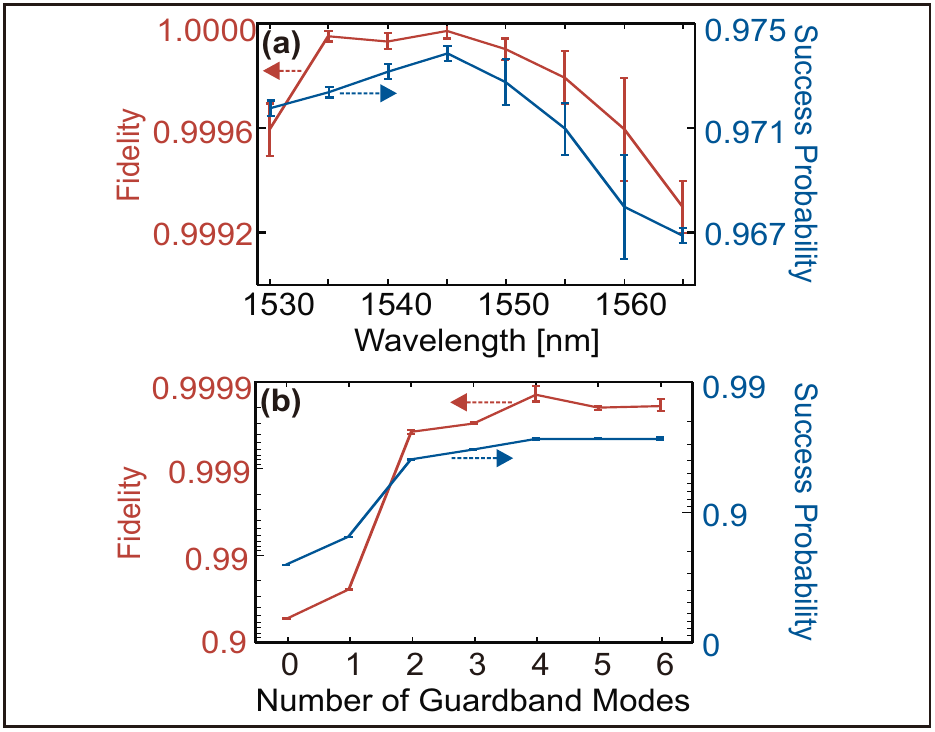}
\caption{(a) Fidelity and success probability as a function of center wavelength. (b) Parallel beamsplitter performance against frequency separation.}
\label{fig3}
\end{figure}

A crucial claim in favor of our beamsplitter is its suitability for parallelization. Ironically, the very characteristic which precludes a deterministic frequency beamsplitter using a single EOM---frequency-translation invariance~(Appendix~\ref{appA})---enables nearly effortless parallelization. After properly compensating dispersion across the optical spectrum (to synchronize group delay between the two EOMs), we scan the wavelength of the central gate mode in 5-nm increments and measure $\mathcal{F}$ and $\mathcal{P}$ at each step over the full C-band. Figure~\ref{fig3}(a) shows that the fidelity exceeds 0.9990 for all test points, and the success probability does not drop below 0.965. A second question, complementary to the total acceptance bandwidth, is the minimum frequency spacing: how close can two single-qubit gates be placed without performance degradation? Since sidebands adjacent to the computational space are populated mid-calculation, one would expect that a finite number of dark, guardband modes are required to prevent cross-contamination. We address this question experimentally by implementing two beamsplitters in parallel and characterizing the total operation as a function of the number of initially empty modes between mode 1 of the low-frequency gate and mode 0 of the higher frequency one. The fidelity and success probability for the collective parallel operation are plotted in Fig.~\ref{fig3}(b); they reach their asymptotic values for separations of just four modes. Combined with the 40-nm (5-THz) bandwidth of Fig.~\ref{fig3}(a) and the 25-GHz mode spacing, these results imply that the present system can realize 33 frequency beamsplitters in parallel---a remarkable indication of the promise of our approach in scalable QIP.

\textit{Frequency tritter.---}Thus far, quantum frequency mixers have focused on the basic two-mode case~\cite{Kobayashi2016, Kobayashi2017, McGuinness2010, Clemmen2016, Joshi2017}, yet the inherent high dimensionality of frequency-bin states makes them well-suited for more complex qudit operations as well. Accordingly, generalizing mode mixers to dimensions beyond $d=2$ represents an important milestone for frequency-based QIP. For $d=3$, the most natural operation is the uniform frequency tritter---the frequency analogue of a $3\times 3$ spatial coupler with equal split ratios~\cite{Zeilinger1993}, which has been shown to enable fundamentally richer quantum physics than the two-mode case~\cite{Menssen2017}. The specificity of such an operation distinguishes the frequency tritter from previous examples of frequency conversion which, while involving three distinct modes, have not attained arbitrary control over the full $3\times 3$ interaction~\cite{Agha2012}. For our purposes, a particularly convenient operation satisfying the equi-amplitude requirement is the 3-point discrete Fourier transform (DFT):
\begin{equation}
\label{e5}
U_{3\times 3} = \frac{1}{\sqrt{3}} \begin{pmatrix} 1 & 1 & 1 \\ 1 & e^{2\pi i/3} & e^{4\pi i/3} \\ 1 & e^{4\pi i/3} & e^{2\pi i/3} \end{pmatrix}.
\end{equation}
Numerically, we find that incorporating an additional harmonic in the EOM drive signals allows our current configuration to reproduce the above frequency tritter with predicted fidelity $\mathcal{F}=0.9999$ and success probability $\mathcal{P}=0.9733$~(Appendix~\ref{appB}). The fact that the modulation  remains so simple even for the tritter operation---consisting of the sum of just two phase-shifted sinewaves---again manifests the fortuitous practicality of our Fourier-series approach, beyond even the original proposal which relied on specialized RF waveforms~\cite{Lukens2017}.

\begin{figure}[!b]
\includegraphics[width=3.4in]{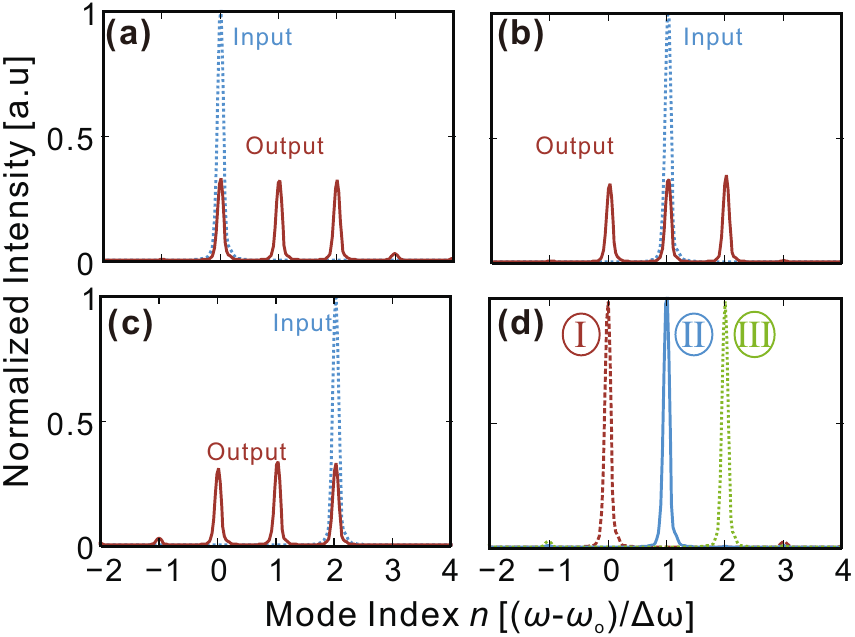}
\caption{Experimentally measured tritter output spectra for specific coherent-state inputs. (a) Pure mode 0:
 $|\alpha_{\omega_0}0_{\omega_1}0_{\omega_2}\rangle$. (b) Pure mode 1: $|0_{\omega_0}\alpha_{\omega_1}0_{\omega_2}\rangle$. (c) Pure mode 2: $|0_{\omega_0}0_{\omega_1}\alpha_{\omega_2}\rangle$. (d) Outputs for the superposition state input $|\alpha_{\omega_0}(e^{-i\phi}\alpha)_{\omega_1}(e^{-2i\phi}\alpha)_{\omega_2}\rangle$ for: (I) $\phi=0$, (II) $\phi=2\pi/3$, and (III) $\phi=4\pi/3$.
 }
\label{fig4}
\end{figure}
Experimentally, we incorporate an RF frequency doubler into the setup (see dotted box in Fig.~\ref{fig1}) to produce the necessary second harmonic. Because of the high-frequency rolloff of our microwave components, we also reduce the drive frequency---and hence, mode spacing---from 25 GHz to 18.1 GHz, for a doubled component at 36.2 GHz~\footnote{Higher frequencies could be obtained by using appropriate V-band (40-75 GHz) hardware}. Running the coherent-state-based characterization algorithm~(Ref. \cite{Rahimi2013} and Appendix~\ref{appD}), we measure fidelity $\mathcal{F} = 0.9989 \pm 0.0004$ and success probability $\mathcal{P} = 0.9730 \pm 0.0002$, again extremely close to theoretical predictions. Figure~\ref{fig4} plots several important input/output spectra: for any single-line input, the output exhibits equal lines in the same three modes; conversely, three-mode input superpositions of the appropriate phases excite single lines at the output. This high-fidelity, balanced frequency tritter---the first of its kind---confirms that our electro-optic technique scales well to higher dimensions, with only a minor increase in the system complexity.

\begin{figure}[!b]
\includegraphics[width=3.4in]{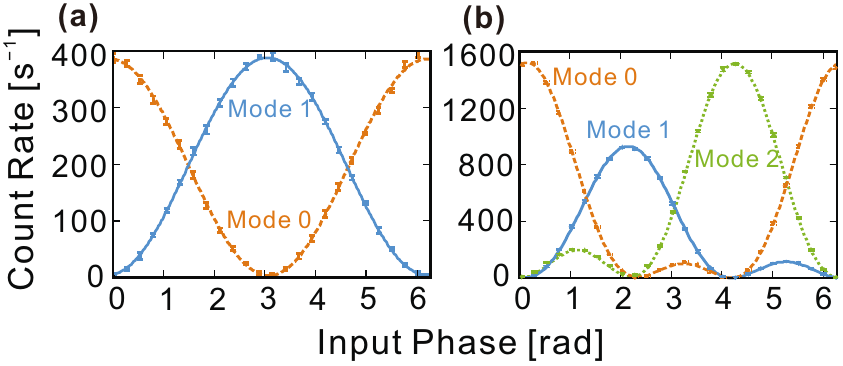}
\caption{Spectral interference with weak coherent states. (a) Output count rates for the two frequency modes of the beamsplitter, as the phase $\phi$ of the single-photon-level state $|\psi_2\rangle$ is scanned. (b) Counts for the three output modes of the frequency tritter as the phase $\phi$ of the three-mode state $|\psi_3\rangle$ is scanned. The plotted best-fit curves are Fourier series of the form $\sum_n A_n \cos(n\phi+B_n)$, summed from $n=0$ to 1 for (a), and $n=0$ to 2 for (b).}
\label{fig5}
\end{figure}

\textit{Single-photon level.---}Finally, to verify that these frequency mode mixers maintain performance at the single-photon level, we attenuate the input state $|\alpha_{\omega_0}(e^{-i\phi}\alpha)_{\omega_1}\rangle$ for the beamsplitter and $|\alpha_{\omega_0}(e^{-i\phi}\alpha)_{\omega_1}(e^{-2i\phi}\alpha)_{\omega_2}\rangle$ for the tritter to $\sim$0.1 photons per detection window at the gate input (i.e., before loss through the frequency mixer) and scan the input phase $\phi$. The resulting interference patterns for these weak coherent states allow us to predict operation fidelity for true single-photon states as well. This follows because the gate itself is a one-photon operation, and thus the interference visibility depends only on the average flux and any extra noise introduced by the gate---not on the photon number statistics of the input. At each setting, we use a wavelength-selective switch to direct the output modes to a gated InGaAs single-photon detector. Figure~\ref{fig5}(a) plots the counts in modes 0 and 1 for the beamsplitter, after subtracting the average detector dark count rate (error bars give the standard deviation of five repeated measurements). Moving on to the three-mode case, we obtain the detection rates for modes 0, 1, and 2 shown in Fig.~\ref{fig5}(b). The oscillations now trace a sum of two sines, with respective peaks at $\phi=0,2\pi/3$, and $4\pi/3$, as expected for the ideal matrix in Eq.~(\ref{e5}). The reduced flux for mode 1 is primarily due to the wavelength-selective switch, as its 12.5-GHz passbands do not match the 18.1-GHz line spacing; in our filter definitions, the center of mode 1 is close to one passband edge, and thereby experiences an additional $\sim$1-dB attenuation. Overall, both the beamsplitter and tritter perform exceptionally well at the single-photon level, with detector-dark-count-subtracted visibilities from 97-100\%. Such low-flux visibilities far exceed those of previous $\chi^{(2)}$ or $\chi^{(3)}$ frequency beamsplitters, which suffer from optical noise generated by the powerful pump fields; our approach inherently contributes no excess noise photons, making it particularly well-suited for quantum applications.

\textit{Discussion.---}A major goal moving forward would be to fully integrate this frequency mixer, using on-chip modulators and pulse shapers---not only for reducing overall footprint but also lowering the current $\sim$12.5-dB insertion loss, due primarily to our use of off-the-shelf telecommunication components. While our system's massive bandwidth could soften the impact of loss in the short term, through parallel replication of a desired operation, the ideal solution would be to reduce the loss altogether by improved engineering. An on-chip EOM with $\sim$1-dB loss has already been demonstrated~\cite{Fan2016}, and an integrated pulse shaper with only $\sim$2-dB loss appears reasonable with established silicon-photonic processes~\cite{Aim2017}. Without a doubt, significant challenges remain to synthesize these capabilities onto a monolithic platform, demanding continued research and as-yet-uncharted technological advances. But the current state of the art nevertheless provides legitimate promise for the development of high-throughput on-chip frequency gates, compatible with on-chip quantum frequency combs~\cite{Grassani2015, Reimer2016, Jaramillo2017, Imany2017, Kues2017}. This integration would be extremely valuable, as the importance of electro-optic mixing has already been demonstrated in off-chip probing of the frequency entanglement of such frequency combs~\cite{Imany2017, Kues2017}. However, these examples used only one EOM and therefore suffered large amounts of scattering outside of the computational space (Appendix~\ref{appA}). By contrast, our multiple-EOM scheme permits inherently efficient true quantum gates---essential for the development of large-scale on-chip frequency QIP systems.

Finally, we note a useful connection between our electro-optic results and previous parametric beamsplitters~\cite{Kobayashi2016, Kobayashi2017, McGuinness2010, Clemmen2016, Joshi2017}. Since our technique excels for tightly spaced modes operated in parallel, whereas nonlinearity-based beamsplitters instead perform well for interband modes spaced beyond typical electro-optic bandwidths, one can envision integrating \emph{both} approaches in the same system: computations can be performed in parallel within dense subbands with our technique, and the resulting photonic states can then be spectrally combined by parametric frequency mixers for further processing. In this way, the advantages of both approaches can be leveraged simultaneously, bringing us one step closer to the full utility of photonic QIP with frequency modes.

\begin{acknowledgments}
We thank W.~R. Ray for use of the optical spectrum analyzer and N. Lingaraju for helpful discussions regarding on-chip photonics. This work was performed in part at Oak Ridge National Laboratory, operated by UT-Battelle for the U.S. Department of Energy under contract no. DE-AC05-00OR22725. Funding was provided by ORNL's Laboratory Directed Research and Development Program and National Science Foundation grant ECCS-1407620.
\end{acknowledgments}

\appendix
\section{Single EOM and scatter probability}
\label{appA}
Consider an EOM driven with phase $\varphi (t)$, assumed periodic at the inverse mode spacing ($T=2\pi/\Delta\omega$). Then the input and output frequency modes are related according to $\hat{a}_m^{(\mathrm{out})} = \sum_n c_{m-n} \hat{a}_n^{(\mathrm{in})}$, where $c_n=(2\pi)^{-1}\int_T dt\, e^{i\varphi(t)} e^{in\Delta\omega t}$ are the Fourier series coefficients of the periodic EOM operation. The ideal EOM is therefore invariant to optical frequency translation; mathematically speaking, the operation is a Toeplitz (diagonal-constant) matrix, with coefficients depending only on the frequency difference between input and output modes. Accordingly, any modulation which succeeds in coupling, say, mode $n$ to $n+1$, will also couple modes $n+1$ to $n+2$ with equal weight. In the case of a uniform mode mixer, this implies that a single EOM will necessarily scatter input photons out of the $d$-mode computational space into adjacent sidebands. Because of the Toeplitz condition, an EOM that attains uniform amplitude for a $d\times d$ matrix must have at least $2d-1$ equal coefficients in its Fourier series (with additional sidebands to preserve unitarity). As $d-1$ of these fall outside of the computational space, the scatter probability is at least $(d-1)/(2d-1)$ for a uniform $d$-mode mixer based on one EOM.

\section{Optimization Approach}
\label{appB}
While our original spectral Hadamard gate makes use of two pairs of pulse shapers and EOMs~\cite{Lukens2017}, we note that three total components (EOM-shaper-EOM) suffice to perform the Hadamard gate with $\mathcal{F}=\mathcal{P}=1$. This follows from the fact that, since the input photon occupies just two frequency modes, any pair of phases applied to these two modes by the first pulse shaper is indistinguishable from a temporal delay: the spectral phase is trivially a linear function of frequency. Thus, any modulation that would have been applied by this pulse shaper can be absorbed into a delay on the first EOM. Under this simplification, we can approximate the transformation matrix for the frequency multiport $V$ by
\begin{equation}
V = F D_3 F^\dagger D_2 F D_1 F^\dagger,
\end{equation}
where $D_1$ and $D_3$ ($D_2$) are diagonal unitary matrices representing the temporal (spectral) phase modulations applied by the EOMs (pulse shaper), and $F$  is the $M \times M$ discrete Fourier transform (DFT). Truncating the matrix $V$ to $M$ modes provides an accurate approximation to the $d\times d$ operation of interest as long as $M\gg d$ and the solution does not experience aliasing---that is, sampling exceeds the Nyquist rate. In our simulations, we utilize the Optimization Toolbox in MATLAB to search for an optimal set of phases for $D_1$, $D_2$, and $D_3$ which preserve fidelity $\mathcal{F} > 0.9999$ and maximize success probability $\mathcal{P}$. In general, each matrix is characterized by $M$ independent real numbers in $(-\pi,\pi]$: for the pulse shaper ($D_2$), these signify the phase shifts applied to each frequency mode; for the EOMs ($D_1$ and $D_3$), these are samples of the temporal phase modulation over one period. However, for experimental practicability, we constrain the temporal phase patterns to sums of sinewaves (i.e., truncated Fourier series), rather than fully arbitrary functions. Thus, taking a total of $p$ harmonics in the optimization---each specified by an amplitude and phase---the number of free parameters for each EOM matrix reduces to $2p$. In the following we set $M=128$ and $p=1$ for the frequency beamsplitter simulations, for a total of $M+2(2p)=132$ numbers to find; for the frequency tritter, we add one more harmonic, giving $p=2$ and 136 total parameters.

\begin{figure}[]
\includegraphics[width=3.4in]{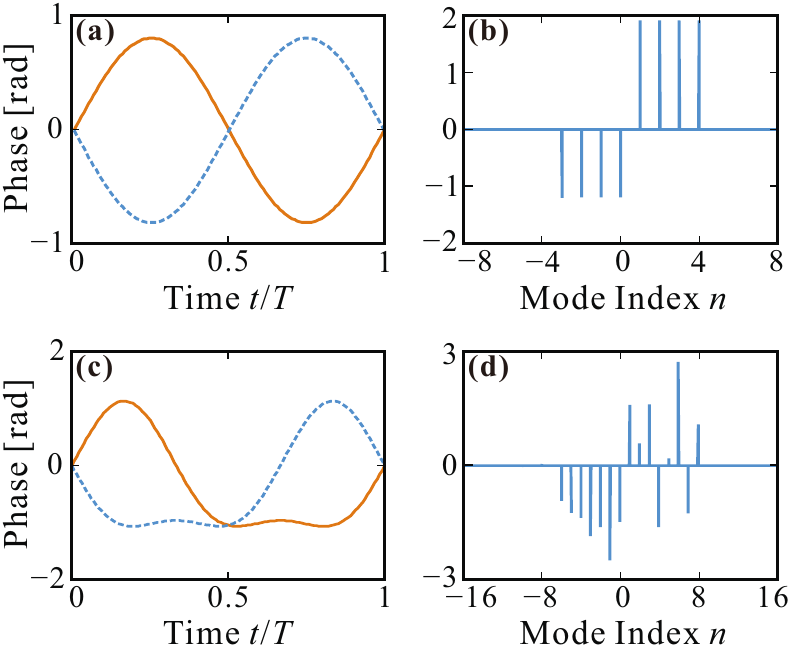}
\caption{Numerical solutions for the time-frequency phases required to implement optimal beamsplitter and tritter. For the frequency beamsplitter: (a) temporal phase modulation applied to the first EOM [solid red] and second EOM [dotted blue], plotted over one period; (b) phases applied to each frequency mode by the pulse shaper, where modes 0 and 1 denote the computational space. For the frequency tritter: (c) temporal phase modulation for first [solid red] and second [dotted blue] EOM; (d) phases applied to each frequency mode by the pulse shaper, where now modes 0, 1, and 2 denote the computational space.}
\label{S1}
\end{figure}

Here we record the specific solutions for the pulse shaper and each EOM in the optimal frequency beamsplitter and tritter. Figures \ref{S1}(a) and (b) show the results for the frequency beamsplitter, with $\mathcal{F} =0.9999$ and $\mathcal{P} =0.9760$. The temporal phases on both EOMs are just phase-shifted sinewaves driven by a single RF tone. In addition, the spectral phase on the pulse shaper, shown in Fig. \ref{S1}(b), turns out to be a step function with a $\pi$-phase jump between mode indices 0 and 1, readily implemented in the line-by-line pulse shaping scheme.

Furthermore, additional simulations show that this three-element setup can implement the frequency DFT up to $d=7$ while maintaining  $\mathcal{F\times P}>0.97$, using drive signals consisting of only $d-1$ single-frequency harmonics~\cite{Untitled}. These findings indicate favorable scaling in our paradigm, effectively sublinear in the number of components and preserving high $\mathcal{F}$ and $\mathcal{P}$.

The solution for the frequency tritter is presented in Figs. \ref{S1}(c) and (d). We incorporate an additional RF harmonic to both EOMs while maintaining the three-element setup, and numerically we achieve $\mathcal{F}=0.9999$ and $\mathcal{P} = 0.9733$. As shown in Fig. \ref{S1}(c), the temporal phases are still time-shifted replicas, but now composed of two harmonics. The introduction of the additional harmonic couples more optical power to high-frequency modes, and relatively more complicated spectral phase control is needed for the frequency tritter, as plotted in Fig. \ref{S1}(d). Note that both solutions are achievable experimentally: the maximum temporal phase shifts [Figs. \ref{S1}(a) and (c)] are well within values available from commercial EOMs, and the number of frequency modes requiring spectral shaping is $\lesssim 20$ [Figs. \ref{S1}(b) and (d)]---much less than the full $M$-mode space, indicating 128 samples are fully sufficient to characterize the solution. This intuition is confirmed numerically; by inserting passbands which block all frequencies beyond a finite interior band, we find no reduction in either $\mathcal{F}$ or $\mathcal{P}$ to 6 significant digits, when keeping just 8 modes for the beamsplitter solution and 16 modes for that of the tritter.

\section{Experimental Methods}
\label{appC}

\subsection{Frequency Beamsplitter}
In our experimental scheme (Fig.~\ref{fig1}), the preparation of input states, frequency mixing, and final output state detection are all built on commercial fiber-optics instrumentation, such as intensity/phase modulators, pulse shapers and single-photon counters. The implementation of the frequency beamsplitter can be described as follows. A tunable continuous-wave (CW) laser operating in the C-band is firstly sent to an intensity modulator (IM; Photline MX-LN-40) driven at 25 GHz, which creates a total number of three frequency bins with a spacing of 25 GHz. (The use of an intensity, rather than phase, modulator was purely from equipment availability: a phase modulator would produce many more comblines with greater efficiency, but the IM suffices for the number of modes needed in this experiment.) The subsequent pulse shaper (Finisar WaveShaper 1000S)---which possesses $\sim$10-GHz spectral resolution, 1-GHz addressability, with operating wavelength from 1527.4 nm to 1567.5 nm---then performs amplitude and phase filtering to prepare either pure mode or superposition states as input to the following frequency beamsplitter.  We use an RF oscillator (Agilent E8257D) to generate a 25-GHz sinewave, and split it three ways feeding amplifiers for the first IM for state preparation and the two 40-Gb/s EOMs (EOSpace) of the frequency beamsplitter. Accurate control of the amplitude and the timing of RF signals is achieved by the usage of variable attenuators and phase shifters, by which we fine tune every RF component until we have correlation above 99.9\% between the experimentally obtained intensity spectrum after each EOM and the theoretical prediction. With estimated EOM half-wave voltages of $V_\pi=5.37$ V at 25 GHz, the total RF power required at each EOM for the solution in Fig.~\ref{S1}(a) is roughly 12.9 dBm.

The central pulse shaper (another Finisar WaveShaper), applies the numerically obtained phase patterns [Fig.~\ref{S1}(b)], and for the parallelization tests (Fig.~\ref{fig3}), it also compensates optical dispersion. Experimentally, we found that applying a dispersion of $-0.4$ ps/nm was sufficient to compensate all frequency-dependent delay between the two EOMs (including the residual dispersion in the pulse shaper itself) and thus ensure proper timing between EOMs across the full C-band. Otherwise, the beasmplitter would not be able to preserve the correct split ratio for all parallel gates \emph{simultaneously}; on the other hand no dispersion compensation is needed for a single gate, since frequency-dependent delay over the bandwidth involved ($\sim$6 modes or $\sim$1 nm) is much smaller than the 40-ps RF period. For output state detection in the high-flux regime, we utilize an optical spectrum analyzer (OSA; Yokogawa) to obtain five spectra for each input state, and calculate the mean and standard deviation for both $\mathcal{F}$ and $\mathcal{P}$. For this coherent-sate characterization, we set the CW laser power to about 5 mW at the gate input.

In the weak-coherent-state experiment, the output state is frequency-demultiplexed by a frequency-selective switch (Finisar WaveShaper 4000S), and measured by an InGaAs single-photon avalanche photodiode (Aurea Technology SPD\_AT\_M2), gated at 1.25 MHz, with a 1-ns gate and 20\% detection efficiency. As the input state is attenuated to $\sim$0.1 photons per detection window at the gate input (inferred by the measured system loss and detector parameters), we register $\sim$400 counts/s on the detector. The dark count rate is measured when the laser is turned off and maximum ($>$35-dB) attenuation is set on the shaper; roughly 20 counts/s are registered. For each phase setting, we perform five 5-s measurements to record mean photon counts and the error bars, subtract the mean dark counts, and calculate the visibility of each trace with the Curve Fitting Toolbox in MATLAB, repeating this for both frequency modes.

\subsection{Frequency Tritter}
For the frequency tritter, we incorporate an RF frequency doubler (Spacek Labs AQ-2X) to produce the necessary second-harmonic signal. Due to a combination of doubling efficiency and loss in current microwave components, we chose for these experiments to operate at 18.1-GHz mode spacing, rather than the beamsplitter's 25 GHz. [No such reduction would be required with all-V-band (40-75 GHz) hardware.] Considering the predicted EOM half-wave voltages at 18.1 and 36.2 GHz ($V_\pi=4.78$ and 6.02 V, respectively), the expected RF power at the input of each EOM is 14.1 dBm at 18.1 GHz and 7.89 dBm at 36.2 GHz. Also, because of the relative difficulty to manually phase shift both harmonics synchronously, we set the relative phase of the two combined frequencies on both EOMs independently, then match the overall delay between EOMs by applying additional linear spectral phase on the central pulse shaper.

The high-flux $\mathcal{F}$ and $\mathcal{P}$ measurements use the same measurement components as in the beamsplitter case. Yet for the single-photon-level tritter tests, demultiplexing is achieved with an amplitude-only wavelength selective switch (Finisar WSS) having 12.5-GHz channel specificity across a total bandwidth of 4.825 THz, and detection with an InGaAs photon counter operated at 4-MHz gate frequency, 2.5-ns gate duration, and $>$10\% efficiency (ID Quantique id-200). Such differences in demultiplexing and detection explain why the overall count rates in the main text (Fig.~5) vary between the beamsplitter and tritter. Measuring dark counts with the same procedure as with the beamsplitter, we obtain $\sim$150 counts/s, which are subsequently subtracted from the totals.

\section{Procedure for Measuring Transformation Matrix}
\label{appD}
Our calculations of $\mathcal{F}$ and $\mathcal{P}$ rely on complete characterization of the $d\times d$ multiport $V_{d\times d}$. We utilize an analogue of the spatial technique shown in Ref.~\cite{Rahimi2013}, and here we provide additional details on precisely how to determine each of the matrix elements. This technique relies on high-power coherent state probing, which is justified because the operation of interest is, at its basic level, a linear multiport; thus its distinguishing behavior holds for high-flux coherent states as well as single photons.

The definition of success probability $\mathcal{P}$ is
\begin{equation}
\mathcal{P} = \frac{1}{d} \Tr (V_{d\times d}^\dagger V_{d\times d}),
\end{equation}
where $V_{d\times d}$ denotes the infinite-dimensional transformation $V$ truncated to the $d$ modes of the desired operation $U_{d\times d}$. This can be written equivalently as
\begin{equation}
\mathcal{P} = \frac{1}{d} \sum_{m=0}^{d-1} \sum_{n=0}^{d-1} |V_{nm}|^2,
\end{equation}
from which we see that $\mathcal{P}$ depends on only the moduli of the $d^2$ matrix elements. To find these values, we probe our frequency multiport with a single optical frequency from index $n=0$ to $d-1$. The information we need to calculate $\mathcal{P}$, namely $|V_{mn}|^2$, is then given by the output optical power in mode $m$ when the input is set to $n$. And by measuring the total throughput of the system in all modes (even those beyond $d$), we can normalize each matrix element by overall transmissivity, distinguishing the insertion loss (photon is missing) from scatter loss (photon remains, but has left $d$-dimensional subspace), so that $\mathcal{P}$ can quantify the latter. Thus, a value $\mathcal{P}=1$ means that, given that the input photon exits the system, it is guaranteed to have undergone the desired operation and has remained in the $d$-mode computational subspace.

On the other hand, the fidelity $\mathcal{F}$ involves the full Hilbert-Schmidt inner product:
\begin{equation}
\mathcal{F} = \frac{1}{d} \frac{|\Tr (V_{d\times d}^\dagger U_{d\times d})|^2}{\Tr (V_{d\times d}^\dagger V_{d\times d})},
\end{equation}
or alternatively
\begin{equation}
\mathcal{F} = \frac{1}{d^2 \mathcal{P}} \left| \sum_{m=0}^{d-1} \sum_{n=0}^{d-1} V_{nm}^* U_{nm} \right|^2,
\end{equation}
which indeed depends on the phase as well as amplitude information of $V_{d\times d}$. To determine these phases, we next probe the setup with superpositions of two frequency modes, scanning the relative phase $\phi$ from $0$ to $2\pi$. Extracting the power on specific modes from a series of optical spectra yields interference patterns over $\phi$, and the unknown phase terms in $V_{d\times d}$ can be obtained by performing sinusoidal fitting on each curve.

In our experiments, we apply the above technique to $d=2$ (beamsplitter) and $d=3$ (tritter). The corresponding frequency multiport matrices are $V_{2\times 2}$ and $V_{3\times 3}$, and the input optical field $E(t)=\sum_{m=0}^{d-1}\sqrt{p_m} e^{i\phi_m} e^{-i\omega_n t}$ can be expressed in mode matrix form as $[\sqrt{p_0} e^{i\phi_0 } \; \sqrt{p_1} e^{i\phi_1} \cdots \sqrt{p_{d-1}} e^{i\phi_{d-1}} ]^T$. We write a general matrix element of $V_{d\times d}$ in polar form as $V_{mn}=r_{mn} e^{i\phi_{mn}}$. Since phase is only physically meaningful up to a unitary rotation, we follow the procedure of Ref.~\cite{Rahimi2013} and define the phases of the first row and column as zero: this effectively provides a reference for zero phase on our input state preparation. Finally, though the matrices in the following equations are expressed in $d$ dimensions for brevity, experimentally the optical power can be scattered out of the $d$-mode computational space into adjacent sidebands. Therefore, the sensitivity of the OSA should be high enough so that we can collect the optical power in as many modes as possible for accurate normalization. Experimentally, we found that only 6-8 modes were needed to encompass all the optical power (to within $10^{-4}$ accuracy).

The test cases for a single-frequency-mode probe are (note that the OSA functions as a frequency-resolved square-law detector):
\begin{textblock}{7}(3.2,8.25)
\fbox{$2\times 2$}
\end{textblock}
\begin{textblock}{7}(3.2,10.15)
\fbox{$3\times 3$}
\end{textblock}

\begin{textblock}{7}(1.1,14.25)
\fbox{$2\times 2$}
\end{textblock}

\begin{widetext}
\centering
\begin{equation*}
\begin{bmatrix} r_{00} & r_{01} \\
r_{10} & r_{11}e^{i\phi_{11}} \end{bmatrix}
\begin{bmatrix} \sqrt{p} \\ 0 \end{bmatrix}
=
\sqrt{p} \begin{bmatrix} r_{00} \\ r_{10} \end{bmatrix}
\xrightarrow[]{\text{OSA}}
p \begin{bmatrix} r_{00}^2 \\ r_{10}^2 \end{bmatrix}
\end{equation*}

\begin{equation*}
\begin{bmatrix} r_{00} & r_{01} \\
r_{10} & r_{11}e^{i\phi_{11}} \end{bmatrix}
\begin{bmatrix} 0 \\ \sqrt{p} \end{bmatrix}
=
\sqrt{p} \begin{bmatrix} r_{01} \\ r_{11}e^{i\phi_{11}} \end{bmatrix}
\xrightarrow[]{\text{OSA}}
p \begin{bmatrix} r_{01}^2 \\ r_{11}^2 \end{bmatrix}
\end{equation*}

\begin{equation*}
\begin{bmatrix}
r_{00} & r_{01} & r_{02} \\
r_{10} & r_{11}e^{i\phi_{11}} & r_{12}e^{i\phi_{12}} \\
r_{20} & r_{21}e^{i\phi_{21}} & r_{22}e^{i\phi_{22}} \\
\end{bmatrix}
\begin{bmatrix}
\sqrt{p} \\ 0 \\ 0 \end{bmatrix}
= \sqrt{p}
\begin{bmatrix}
r_{00} \\ r_{10} \\ r_{20} \end{bmatrix}
\xrightarrow[]{\text{OSA}}
p
\begin{bmatrix}
r_{00}^2 \\
r_{10}^2 \\
r_{20}^2
\end{bmatrix}
\end{equation*}

\begin{equation*}
\begin{bmatrix}
r_{00} & r_{01} & r_{02} \\
r_{10} & r_{11}e^{i\phi_{11}} & r_{12}e^{i\phi_{12}} \\
r_{20} & r_{21}e^{i\phi_{21}} & r_{22}e^{i\phi_{22}} \\
\end{bmatrix}
\begin{bmatrix}
0 \\
\sqrt{p} \\
0
\end{bmatrix}
= \sqrt{p}
\begin{bmatrix}
r_{01} \\
r_{11}e^{i\phi_{11}} \\
r_{21}e^{i\phi_{21}}
\end{bmatrix}
\xrightarrow[]{\text{OSA}}
p
\begin{bmatrix}
r_{01}^2 \\
r_{11}^2 \\
r_{21}^2
\end{bmatrix}
\end{equation*}

\begin{equation}
\begin{bmatrix}
r_{00} & r_{01} & r_{02} \\
r_{10} & r_{11}e^{i\phi_{11}} & r_{12}e^{i\phi_{12}} \\
r_{20} & r_{21}e^{i\phi_{21}} & r_{22}e^{i\phi_{22}} \\
\end{bmatrix}
\begin{bmatrix}
0 \\
0 \\
\sqrt{p}
\end{bmatrix}
= \sqrt{p}
\begin{bmatrix}
r_{02} \\
r_{12}e^{i\phi_{12}} \\
r_{22}e^{i\phi_{22}}
\end{bmatrix}
\xrightarrow[]{\text{OSA}}
p
\begin{bmatrix}
r_{02}^2 \\
r_{12}^2 \\
r_{22}^2
\end{bmatrix}
\end{equation}
\justify
We thus see that by these measurements we can obtain all $d^2$ amplitudes of $V_{d\times d}$. Subsequently, we probe the system with superpositions of two frequency modes, and scan the relative phase $\phi \in [0,2\pi]$  between them. The different configurations are:

\centering
\begin{equation*}
\begin{bmatrix} r_{00} & r_{01} \\
r_{10} & r_{11}e^{i\phi_{11}} \end{bmatrix}
\begin{bmatrix} \sqrt{p} \\ \sqrt{p}e^{i\phi}  \end{bmatrix}
=
\sqrt{p} \begin{bmatrix} r_{00}+r_{01} e^{i\phi} \\ r_{10}+r_{11} e^{i(\phi+ \phi_{11})} \end{bmatrix}
\xrightarrow[]{\text{OSA}}
p \begin{bmatrix} r_{00}^2 + r_{01}^2 + 2r_{00}r_{01} \cos\phi \\
r_{10}^2 + r_{11}^2 + 2 r_{10}r_{11} \cos(\phi+\phi_{11}) \end{bmatrix}
\end{equation*}

\begin{equation*}
\begin{bmatrix}
r_{00} & r_{01} & r_{02} \\
r_{10} & r_{11}e^{i\phi_{11}} & r_{12}e^{i\phi_{12}} \\
r_{20} & r_{21}e^{i\phi_{21}} & r_{22}e^{i\phi_{22}} \\
\end{bmatrix}
\begin{bmatrix} \sqrt{p} \\ \sqrt{p}e^{i\phi} \\ 0  \end{bmatrix}
=
\sqrt{p} \begin{bmatrix} r_{00}+r_{01} e^{i\phi} \\ r_{10}+r_{11} e^{i(\phi+ \phi_{11})} \\  r_{20}+r_{21} e^{i(\phi+ \phi_{21})} \end{bmatrix}
\xrightarrow[]{\text{OSA}}
p \begin{bmatrix} r_{00}^2 + r_{01}^2 + 2r_{00}r_{01} \cos\phi \\
r_{10}^2 + r_{11}^2 + 2 r_{10}r_{11} \cos(\phi+\phi_{11}) \\
r_{20}^2 + r_{21}^2 + 2 r_{20}r_{21} \cos(\phi+\phi_{21})
\end{bmatrix}
\end{equation*}

\begin{textblock}{7}(-1.75,1.7)
\fbox{$3\times 3$}
\end{textblock}

\begin{equation}
\begin{bmatrix}
r_{00} & r_{01} & r_{02} \\
r_{10} & r_{11}e^{i\phi_{11}} & r_{12}e^{i\phi_{12}} \\
r_{20} & r_{21}e^{i\phi_{21}} & r_{22}e^{i\phi_{22}} \\
\end{bmatrix}
\begin{bmatrix} \sqrt{p} \\ 0 \\ \sqrt{p}e^{i\phi}   \end{bmatrix}
=
\sqrt{p} \begin{bmatrix} r_{00}+r_{02} e^{i\phi} \\ r_{10}+r_{12} e^{i(\phi+ \phi_{12})} \\  r_{20}+r_{22} e^{i(\phi+ \phi_{22})} \end{bmatrix}
\xrightarrow[]{\text{OSA}}
p \begin{bmatrix} r_{00}^2 + r_{02}^2 + 2r_{00}r_{02} \cos\phi \\
r_{10}^2 + r_{12}^2 + 2 r_{10}r_{12} \cos(\phi+\phi_{12}) \\
r_{20}^2 + r_{22}^2 + 2 r_{20}r_{22} \cos(\phi+\phi_{22})
\end{bmatrix}
\end{equation}
\end{widetext}
For each curve, we then perform sinusoidal fitting with respect to the input phase $\phi$ and obtain all the phase values in $V_{d\times d}$. And from this, we can calculate fidelity $\mathcal{F}$.

To give an idea of what our measurements produce, we provide two examples of matrices obtained using the previous characterization method. An example mode transformation for the beamsplitter is
\begin{equation}
V_{2\times 2} = \begin{bmatrix} \sqrt{0.4871} & \sqrt{0.4869} \\
\sqrt{0.4866} & \sqrt{0.4871}e^{i3.1400} \end{bmatrix}.
\end{equation}
These values correspond to $\mathcal{P}=0.9739$ and $\mathcal{F}=0.9999$ when compared to the ideal Hadamard gate. Error bars from repeating the full characterization four more times then gave $\mathcal{P} = 0.9739 \pm 0.0003$ and $\mathcal{F} = 0.99998 \pm 0.00003$.

For the three-mode DFT, an example transformation measured is
\begin{equation}
V_{3\times 3} = \begin{bmatrix} \sqrt{0.3261} & \sqrt{0.3126} & \sqrt{0.3062} \\
\sqrt{0.3183} & \sqrt{0.3290}e^{i2.0925} & \sqrt{0.3339}e^{i4.1775} \\
\sqrt{0.3202} & \sqrt{0.3476}e^{i4.1365} & \sqrt{0.3256}e^{i2.0425}
\end{bmatrix},
\vspace{0.15in}
\end{equation}
with associated success $\mathcal{P}=0.9731$ and fidelity $\mathcal{F}=0.9992$ with respect to the perfect (i.e., not numerically simulated) DFT matrix. Averaging over five repeated measurements then yielded  $\mathcal{P} = 0.9730 \pm 0.0002$ and $\mathcal{F} = 0.9989 \pm 0.0004$, as in the main text.

%

\end{document}